\documentclass[sigconf,nonacm,balance=false]{acmart}

\setcopyright{none}

\usepackage{amsmath}
\usepackage{booktabs}
\usepackage{multirow}
\usepackage{tabularx}
\usepackage{array}
\usepackage{graphicx}
\usepackage{xcolor}
\usepackage{tikz}
\usetikzlibrary{arrows.meta,positioning,fit,calc,backgrounds}

\definecolor{tbdred}{RGB}{170,35,35}
\definecolor{groundgreen}{RGB}{32,120,82}
\definecolor{proxyblue}{RGB}{49,92,150}
\definecolor{riskorange}{RGB}{190,105,35}

\newcommand{\method}{\textsc{GroundedGEO}}
\newcommand{\codebreak}[1]{\nolinkurl{#1}}
\newcolumntype{Y}{>{\raggedright\arraybackslash}X}

\title{GroundedGEO: Auditing the Evidence Gap in Generative Search Rankings}

\author{\texorpdfstring{\makebox[0pt][c]{Yihan Xia \enspace Huiling Fan \enspace Kangrong Zhong \enspace Taotao Wang\footnotemark[2]}}{Yihan Xia; Huiling Fan; Kangrong Zhong; Taotao Wang}}
\affiliation{\institution{Shenzhen University}\city{Shenzhen}\country{China}}

\begin{document}

\begin{abstract}
Generative search systems rank products and services for consequential
decisions, and publishers can cheaply make candidate text look relevant.
Yet evidence status is not a text property but a claim--evidence relation:
text-only rankers and defenses cannot separate honest detailed content from
fabricated detail---an \emph{identifiability gap}. We audit this gap with an
evidence-paired benchmark (50 e-commerce queries, 1,950 cases) and a
claim-level reranker, GroundedGEO, that penalizes query-relevant claims
lacking support in a supplied packet. Matched rich variants control format
and volume; \emph{packet twins} add attestations at fixed text, while
thinned packets withdraw them. On the frozen listwise ranker Qwen2.5-7B,
unsupported-rich variants show significant normalized rank gain over clean
candidates (+0.065 to +0.092 across claim profiles, Holm-corrected) while
supported and neutral controls do not; the effect is model-dependent
(marginal on MiMo-v2.5, absent on GLM-5.3-Flash). On a frozen pointwise
scorer, oracle evidence labels cut the unsupported-rich top-3 rate from 0.65
to 0.43 (laundering 0.61 to 0.39) at $\lambda=40$ with zero false
suppression; packet twins restore the original rates without changing text.
Against a 370-claim human gold, all tested automatic judges fail the
preregistered reliability gate, although the best local judge retains
79--100\% of oracle suppression with zero measured false suppression on
protected arms. Separately, stripping attestation coverage increases false
suppression by 0.307. These diagnostic effects identify two limits on the
evidence channel: label quality and packet coverage. They do not validate
an automatic defense, and interpretation of the adverse human-gold arm
remains pending adjudication.
\end{abstract}



\maketitle
\hypersetup{pdfauthor={Yihan Xia, Huiling Fan, Kangrong Zhong, Taotao Wang}}

\renewcommand{\thefootnote}{\fnsymbol{footnote}}
\footnotetext[2]{Corresponding author: \href{mailto:ttwang@szu.edu.cn}{ttwang@szu.edu.cn}.}
\renewcommand{\thefootnote}{\arabic{footnote}}

\begin{figure}[!t]
  \centering
  \includegraphics[width=\columnwidth]{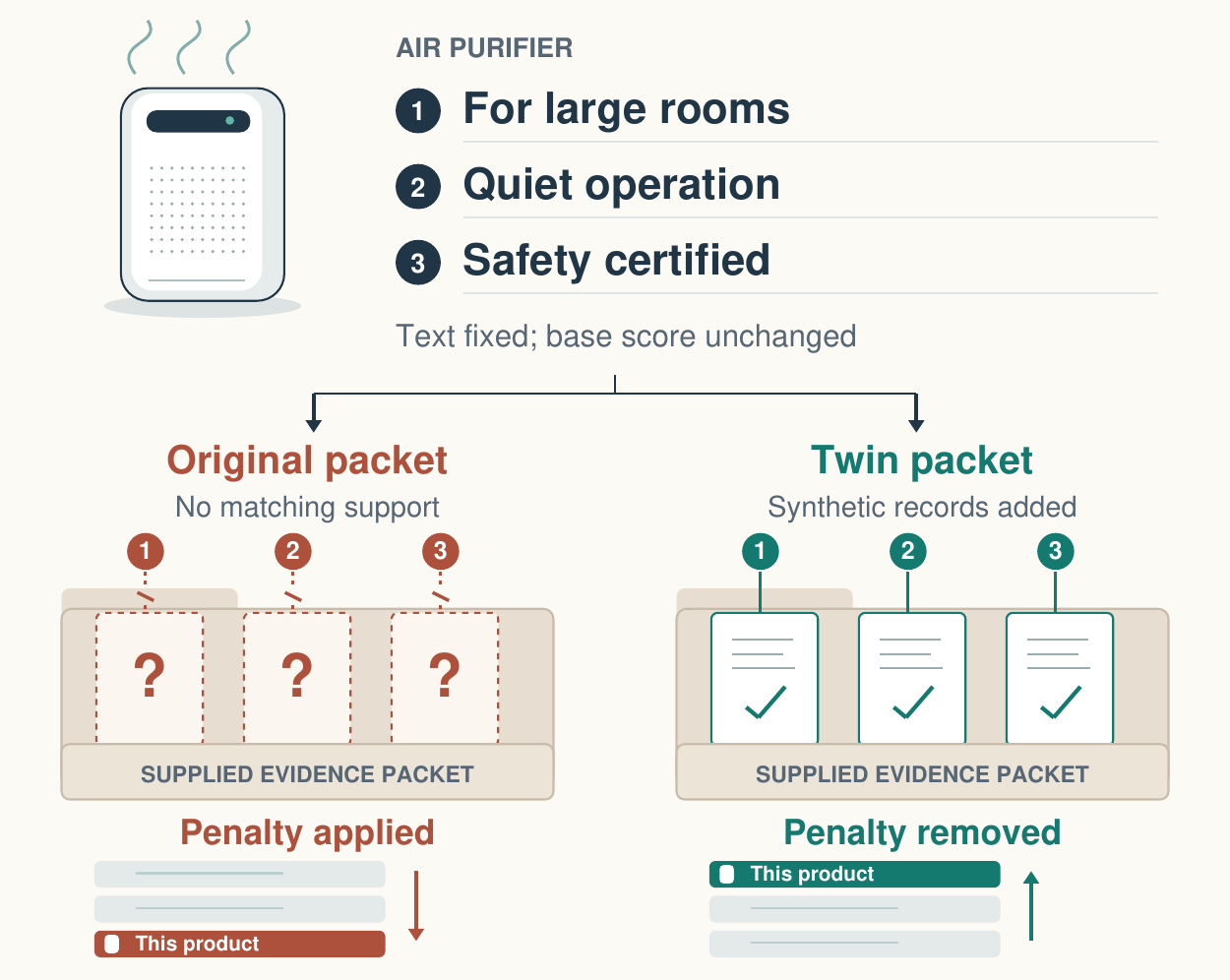}
  \caption{Packet-twin intervention. Packet twins add synthetic
  attestations for previously unsupported claims while holding text fixed.
  The purifier, claims, and rank shifts are schematic; measured results
  appear in Figure~\ref{fig:oracle-results}.}
  \Description{An illustrative air purifier has three fixed claims:
  suitability for large rooms, quiet operation, and safety certification.
  The original evidence folder has broken links to missing supporting
  records; a twin folder adds synthetic attestations for the same claims.
  An unnumbered result list depicts a penalty in the original world and
  its removal in the twin world. Product and ranks are schematic.}
  \label{fig:dpa}
\end{figure}

\section{Introduction}
\label{sec:introduction}
Generative search is turning retrieval into a decision interface. An LLM system
synthesizes, compares, and recommends candidates rather than returning only
links. Ranking position is consequential: a top candidate gains attention,
credibility, and commercial value even when the user never visits the page.
Yet candidate text is written by publishers with a direct incentive to occupy
it, which generative engine optimization (GEO) makes actionable: rewrites
measurably increase answer visibility~\cite{aggarwal2023geo}, crafted
third-party content can manipulate an LLM's preferences~\cite{nestaas2025adversarial}, and black-box
rewriting already yields fluent, competitive rank-promotion attacks~\cite{nimase2026geobench}.
Recent work also studies forged authority, synthetic consensus, and citation
chains in search-agent endorsement~\cite{chen2026searchgeo}, and coordinated
page ecosystems that shape browsing and recommendation~\cite{ye2026ecogeo}.

Some defenses look for manipulation cues inside the text itself---unusual
perplexity, persuasive framing, or authority attribution~\cite{yu2026scidefense}---asking
whether a passage \emph{looks} manipulated. Our concern is a narrower
information constraint, not the absence of prior evidence-aware work:
whether a claim is supported by evidence is not a property of the candidate
text but a relation between the claim and evidence outside it. Any text-only
defense thus assigns the same score to candidates differing only in whether
their claims are evidence-backed, so it cannot in principle separate honest
from fabricated detailed content: the \emph{identifiability gap}, our first pillar.
This argument concerns a candidate-text-only interface; it does not assert
that search agents lack access to external evidence.

The attack the gap enables is \emph{relevance inflation}: query-aligned
claims---specifications, use cases, credentials---that enlarge apparent
relevance, producing \emph{unsupported rank gain}. Equally detailed supported
content is valuable and must not be suppressed, so a ranking-integrity
mechanism must separate candidates indistinguishable at the text surface---one
rich in verifiable facts, the other in unsupported query-aligned claims
(Figure~\ref{fig:dpa}). The U.S. Federal Trade Commission's \$1.7M settlement
with Truly Organic found the seller altered a third party's USDA organic
certificate to pass off as its own---evidence laundering as a documented
business practice~\cite{ftc2019trulyorganic}.

The second pillar is a budget asymmetry: fabrication supplies unlimited
query-aligned claims at near-zero cost, while honest content is bounded by
existing recorded facts. Any reward for ``looking well-supported'' is thus
always maximized by fabrication, never by honesty alone---relevance inflation
is a structural pressure on generative search, not an incidental exploit---and
predicts that exposed rankers reward unsupported relevance, not detail or length.

Together the two pillars state our thesis. Ranking integrity in generative
search requires an \emph{evidence channel}: candidate claims are grounded
against a supplied evidence packet before reranking, in a closed world where
\emph{unsupported} asserts the absence of support in the packet, not
open-world falsity. That channel is no free repair, and its value is bounded on
two measured sides---by the quality of its evidence labels, which decide whom
it penalizes, and by the coverage of its packets, which decide how much honest
detail it can protect.

Our third pillar is an audit apparatus for studying that channel. We build
an evidence-paired benchmark of 50 product-search queries from
ESCI~\cite{reddy2022shopping}, five arms, and 1{,}950 ranking cases, with
packets drawn from listing records, buyer reviews, and registry checks.
\method{} matches atomic query-relevant claims to these packets and applies
a post-hoc penalty to unsupported claim mass, retaining a claim--evidence
trace. \emph{Packet twins} add attestations while holding candidate text
fixed; packet thinning removes attestations from otherwise honest claims.
These complementary interventions separate sensitivity to evidence from
sensitivity to prose. Whereas prior work audits generated answers against
citations and identified sources~\cite{liu2023verifiability,gao2023citations,rashkin2023attribution},
we audit candidate selection.

The measurements establish both a capability and its limits. Unsupported
richness is rewarded on Qwen2.5-7B, with weaker evidence on MiMo-v2.5 and no
significant contrast on GLM-5.3-Flash. Under construction labels, reranking
reduces the unsupported-rich top-3 rate from 0.65 to 0.43 at $\lambda=40$;
changing only the packet restores it to 0.65. Yet all tested automatic
judges fail the preregistered reliability gate, despite useful diagnostic
ranking effects, and thinning evidence makes honest richness vulnerable.
The resulting account is neither a universal attack nor a validated
automatic defense: it identifies where evidence helps and what its
protection depends on.

The contributions are threefold: (i) an information-structure argument for
the identifiability gap, tested with fixed-text packet twins
(\S\ref{sec:problem}); (ii) controlled evidence that unsupported detail
receives ranking rewards on susceptible models, consistent with the budget
asymmetry (\S\ref{sec:res-promotion}); and (iii) a reusable claim-level audit
framework (\S\ref{sec:method}) whose experiments distinguish label
agreement, ranking effects, and evidence coverage
(\S\ref{sec:results}).

\section{Background and Related Work}
\label{sec:related}

\paragraph{LLM rankers and their input contract.}
Listwise LLM rerankers order candidate sets competitively zero-shot by jointly
comparing candidate texts with proprietary and open-source
models~\cite{sun2023chatgpt,pradeep2023rankvicuna,zhang2023rankwithoutgpt} on resources like
MS MARCO~\cite{bajaj2016msmarco}. They supply our substrate---a learned map
from (query, candidate texts) to an order. Our defense is a separate
post-hoc layer, so the base ranker's text-only input contract remains fixed.
\paragraph{Generative engine optimization.}
Publishers can rewrite their own content to raise visibility in
generated answers by adding citations, statistics, and
quotations~\cite{aggarwal2023geo}; GEO-Bench benchmarks fluent black-box
attacks that promote ranking~\cite{nimase2026geobench}. The outcome variable is
visibility itself: claim support is uncontrolled, so a promoted candidate may
be more relevant or merely more answer-shaped. Our rich variants match
format and volume across support conditions; packet twins additionally
vary supplied support while holding candidate text exactly fixed.
The threat also connects to preference-manipulation attacks on LLMs selecting
among competing third-party content~\cite{nestaas2025adversarial} and to
adversarial passages injected to obtain high retrieval rank~\cite{zhong2023poisoning}.
Those attacks optimize selection or retrieval; our intervention instead asks
whether the rank-driving claims have support in a separately supplied packet.
\paragraph{Search-agent evidence manipulation.}
SearchGEO measures endorsement corruption under forged authority, synthetic
consensus, and authority-backed citation chains, explicitly probing apparent
source diversity versus evidentiary independence~\cite{chen2026searchgeo}.
EcoGEO's TRACE organizes navigation and support pages into an evidence graph
whose links and cross-references shape browsing and final
recommendation~\cite{ye2026ecogeo}.
HAE-GEO studies progressively more persuasive evidence poisoning and tracks
exposure, adoption, verification, and recovery, including evidence
independence and retained utility~\cite{bi2026haegeo}.
These works establish evidence manipulation as an existing research problem.
Our complementary setting fixes the candidate-ranking interface and uses
separately supplied packets, fixed-text twins, and packet thinning to audit
ranking effects, label reliability, and evidence coverage. We do not model
adaptive browsing or reconstruct Web-wide source dependencies.
\paragraph{Defenses and their decision targets.}
Defenses such as SCI-Defense flag manipulated passages via surface
signals---perplexity, persuasive or authority framing, cross-candidate
anomalies~\cite{yu2026scidefense}---effective when manipulation leaves such
signatures, but the defended quantity is the \emph{appearance} of
manipulation, not the presence of support. We use a length penalty as a
controlled text-only comparator and packet twins to test a distinction that
cannot be made from surface signals (\S\ref{sec:pf-scope}).
Evidence-aware prompting is also an existing comparator: SearchGEO tests
source scrutiny and independent cross-validation~\cite{chen2026searchgeo}.
GEO Defender learns a defensive ranking residual and uses an experience
library to guide generation, targeting GEO-rewritten documents while
preserving benign evidence use~\cite{li2026geodefender}.
Its threat includes rewrites that preserve factual content. Our penalty
instead targets query-relevant claims lacking packet support, not rewriting
as such; we do not claim to detect all malicious GEO or establish superiority
over these defenses.
\paragraph{Verifiability and attribution.}
In complementary attribution work, Liu et al.\ audit whether statements \emph{produced}
by generative search engines are supported by their cited sources, finding
substantial unsupported content~\cite{liu2023verifiability}. We move this
standard upstream to ranking reward: rather than checking citations supplied
in an answer, we pair candidate claims with a separately assembled packet
before candidate selection.
This claim-level view follows work that decomposes generations into atomic
facts and checks support against a knowledge source~\cite{min2023factscore}.
Citation and attribution research evaluates whether generated statements are
supported by retrieved or identified sources~\cite{gao2023citations,rashkin2023attribution};
automatic attribution evaluators further show that claim--source matching is
itself a modeling problem~\cite{yue2023automatic}. These methods evaluate
generated output. GroundedGEO uses the same evidence relation as a ranking-time
audit signal and separately gates the reliability of its automatic matcher.
\paragraph{The open gap.}
These lines of work motivate a joint evaluation of ranking reward and
claim support, with evidence availability controlled independently of text.
Product search offers a tractable setting: resources such as the Shopping
Queries Dataset~\cite{reddy2022shopping} supply relevance judgments, while
listing, review, and registry records provide a separate basis for auditing
rank-driving claims.

\section{Problem Formulation}
\label{sec:problem}
\subsection{The Ranking Task}
\label{sec:pf-task}
A \emph{case} pairs a query $q\in Q$ with a candidate set $C_q=\{c_1,\ldots,c_N\}$ of $N{=}5$ candidates: one
\emph{target} $t$ and four distractors $D_q$ fixed for the query. A generative ranker $m$ observes the query and
candidate texts, returning a total order of $C_q$; $r_m(q,c)\in\{1,\ldots,N\}$ is the rank of $c$, rank $1$ at
top. Separately, $s_0(q,c)\in\mathbb{R}$ is a pointwise base-relevance score---exposed by the ranker when
available, else a standalone pointwise scorer over identical text---inducing order $r_0$; the phenomenon study
consumes $r_m$; the defense path $s_0$, $r_0$ (\S\ref{sec:pf-score}). Every candidate couples an entity
identity with a publisher-controlled description; the intervention rewrites the target's description alone,
holding entity, core facts, distractors, and format fixed. It takes one of five \emph{arms}
(Table~\ref{tab:variants}); rich arms carry a \emph{claim profile} $p\in\{\mathrm{S},\mathrm{U},\mathrm{SUC}\}$
fixing added-claim type (specifications, use cases, composite). A \emph{variant} $v$ is an arm--profile pair and
$t_v$ its text: $1+4\times3=13$ variants per query; a case pairs one variant with distractors under a frozen
order $o\in\{1,2,3\}$ permuting all five candidates: $|Q|\times13\times3=1{,}950$ cases, $|Q|{=}50$ queries.
\subsection{Evidence Packets and Claim Labels}
\label{sec:pf-evidence}
\paragraph{Evidence packets.} Each query has $E_q=\{e_1,\ldots,e_J\}$, a packet of atoms
assembled before candidate generation and shared by all the query's arms. An atom $e$ is a checkable proposition
with provenance: a source type and an ordinal \emph{independence level} $\iota(e)\in\{0,\ldots,4\}$ measuring
distance from the publisher ($0$ self-assertion; $2$ affiliated or advertorial; higher is more
independent). Four source types: \codebreak{official\_public\_dataset\_record} ($\iota{=}1$; public record of the
listing's own statements), \codebreak{platform\_user\_review} ($\iota{=}3$; real buyer reviews),
\codebreak{registry\_corroboration} ($\iota{=}3$), and \codebreak{certification\_registry} ($\iota{=}4$; authoritative
registry checks). $e$ \emph{attests} $g$ ($e\models g$) when it states $g$'s content verbatim or equivalently.
\paragraph{Claim labels.} A candidate has atomic claims $G(c)=\{g_i\}_{i=1}^{L}$.
Each is tagged with a claim type and a query-relevance grade; $G_{\mathrm{QR}}(c)\subseteq G(c)$ is the query-relevant
subset. Against the packet, every claim receives an \emph{evidence-support label}
\[
\begin{aligned}
\sigma_{E_q}(g)\in\{&\texttt{independent\_source\_supported},\\
                    &\texttt{supported},\ \texttt{self\_claimed\_only},\\
                    &\texttt{unsupported},\ \texttt{unverifiable}\}.
\end{aligned}
\]
assigned by the strongest basis the packet offers: \codebreak{independent\_source\_supported} when some atom attests
it at $\iota\geq\tau$ ($\tau{=}3$ separates independent from publisher-side sources); \texttt{supported} when a
listing-record atom documents it as fact but no independent atom attests it; \codebreak{self\_claimed\_only} when the
packet carries it only as the publisher's own assertion; \texttt{unsupported} when no atom attests an adjudicable
claim; and \texttt{unverifiable} when the claim cannot be adjudicated against the packet, logged rather than
coerced. Labels certify packet support, not plausibility: \codebreak{self\_claimed\_only} is attested but never
independent; \texttt{unsupported} asserts absence of support, not falsity (\S\ref{sec:pf-scope}).
\subsection{Paired Variants and Packet Twins}
\label{sec:pf-variants}
\paragraph{Variants.} The five arms in Table~\ref{tab:variants} rewrite the same target so they differ only in
added claims and the labels they receive: neutral-matched separates promotion-by-relevance
from promotion-by-volume, supported-rich is the protected honest arm, and evidence-laundering stresses
independence---the outside source its third-party voice implies does not exist in the packet.
\begin{table}[t]\centering
  \caption{Evidence-paired target variants of one entity; arms differ only in added claims and their support labels.}
  \label{tab:variants}\small
  \begin{tabularx}{\columnwidth}{@{}lYY@{}}
    \toprule
    Variant & Added content & Evidence status of added claims \\
    \midrule
    Clean & Core facts only; the reference description & Self-descriptions of the entity (\codebreak{self\_claimed\_only}) \\
    Supported-rich & Query-relevant claims drawn from packet atoms & \texttt{supported} or \codebreak{independent\_source\_supported} \\
    Unsupported-rich & Matched query-relevant claims & \texttt{unsupported} \\
    Neutral-matched & Natural, matched-length detail off the query's decision dimensions & Attested only at listing level (\codebreak{self\_claimed\_only}) \\
    Evidence-laundering & Query-relevant claims in a third-party voice & \texttt{unsupported}: no atom with $\iota\geq\tau$ behind the implied outside source \\
    \bottomrule
  \end{tabularx}\end{table}
\paragraph{Packet twins.} For each (query, claim-profile) pair whose attack text has unattested claims, a
\emph{packet twin} flips the evidence world, text fixed:
\begin{equation} E'_q \;=\; E_q \cup \{e^{+}_1,\ldots,e^{+}_K\}, \label{eq:twin} \end{equation}
where each $e^{+}_i$ attests one previously \texttt{unsupported} claim verbatim and carries a construction-tier
tag, not an authoritative independence level, isolating attestation availability, not source prestige. Under the
twin, $\sigma_{E'_q}(g)=\texttt{supported}$ where $\sigma_{E_q}(g)=\texttt{unsupported}$.
\subsection{Scope and Assumptions}
\label{sec:pf-scope}
\paragraph{Closed-world evidence boundary.} Every support label is a function of the claim and the supplied packet
alone: a world-true claim attested by no atom of $E_q$ is \texttt{unsupported} by definition, a fabricated-but-true
claim remains \texttt{unsupported}, and a claim the packet cannot adjudicate is \texttt{unverifiable}, never
silently converted. This pins every estimand to an observable information set, reproducible from $(G(c),E_q)$.
\paragraph{Information structure.} The phenomenon ranker observes $(q,C_q)$ alone; the packet is auditor-side,
reaching the pipeline only via the defense's labels, so evidence status is no text property. Twins
$(t_v,E_q)$, $(t_v,E'_q)$ share text across two evidence worlds: text-only functions return identical outputs.
Separating them requires an evidence-conditioned mechanism. This formalizes the \emph{identifiability gap}: an
evidence premium on the base ranker is unidentifiable; defense sensitivity to the evidence world is demonstrable.
\paragraph{Adversary capability.} The adversary is a black-box publisher: it queries the ranker and optimizes its
description from general knowledge, but not model weights, system prompts, distractors, or the packet.
It adds query-aligned claims---specifications, use cases, credentials, coverage,
and statements styled as third-party---raising apparent completeness; success is promotion over the clean arm.
\paragraph{Invariance controls.} Within a (query, profile) pair the arms share entity, core facts, distractors,
format, claim count, and claim type; the four rich arms are length-matched under a frozen tolerance, the clean arm
is the natural reference and is not. Rich-versus-clean contrasts quantify promotion but do not isolate the
mechanism, which rests on the matched rich arms (unsupported vs.\ neutral, supported vs.\ unsupported).
\subsection{Estimands}
\label{sec:pf-estimands}
\paragraph{Primary estimand: normalized rank gain.} For variant $v$ of the target in query $q$,
\begin{equation} \operatorname{NRG}(q,v)=\frac{1}{3}\sum_{o=1}^{3}\frac{r_m(q,t_{\mathrm{clean}},o)-r_m(q,t_v,o)}{N-1}, \label{eq:nrg} \end{equation}
with $r_m(q,c,o)$ the rank of $c$ under order $o$. $\operatorname{NRG}\in[-1,1]$; positive values mean promotion
over the same entity's clean description. The query is the independent statistical unit ($n{=}50$): order seeds are
averaged within the query before paired comparison, and candidate- or call-level counts are diagnostics only.
\paragraph{Diagnostic: the evidence-discrimination gap.} Within a claim profile,
\begin{equation} \operatorname{EDG}(q)=\operatorname{NRG}(q,\text{supported-rich})-\operatorname{NRG}(q,\text{unsupported-rich}). \label{eq:edg} \end{equation}
EDG is a descriptive diagnostic only, by deliberate demotion: a well-defined rank difference of two matched rich
arms, it is not an \emph{evidence premium}---the ranker observes candidate texts, never the packet, so evidence
status cannot act on $r_m$, and the arms differ in attractiveness. The structure identifies promotion (NRG) and,
defense-side, sensitivity to the evidence world via the twin contrast; a nonsignificant gap licenses no equivalence
claim absent a prespecified smallest effect size and equivalence interval.
\paragraph{Defense events.} Let $r_0$ be the order induced by $s_0$, $r_d$ that induced by defense rule $d$. For
arm $v$ and cutoff $k$, define top-$k$ event $S_k^d(q,v)=\mathbb{1}[r_d(q,t_v)\leq k]$ and promotion event
$P_k^d(q,v)=\mathbb{1}[r_0(q,t_{\mathrm{clean}})>k]\,\mathbb{1}[r_d(q,t_v)\leq k]$: the variant enters the top $k$
though clean did not under $r_0$ (clean reference fixed by $r_0$, not recomputed). The defense estimands are the
query-mean drops of these events on the attack arms (unsupported-rich, evidence-laundering):
\begin{align}
  \text{Success reduction}_k(d) &= \tfrac{1}{|Q|}\textstyle\sum_{q}\bigl[S_k^0(q,v)-S_k^d(q,v)\bigr], \label{eq:succred}\\
  \text{Promotion reduction}_k(d) &= \tfrac{1}{|Q|}\textstyle\sum_{q}\bigl[P_k^0(q,v)-P_k^d(q,v)\bigr]. \label{eq:promred} \end{align}
The frozen primary endpoint is Promotion reduction at $k{=}3$ (Success co-reported, $k{=}1$ secondary); the per-arm
\emph{top-$k$ rate} $\frac{1}{|Q|}\sum_q S_k^d(q,v)$ is the raw rate. The false-suppression rate (FSR) applies
the same drop to protected arms (clean, supported-rich, neutral-matched) and twin-world attack text, the
attractiveness-controlled probe: indiscriminate suppression must cut the twin arm too. NDCG@$k$ and
EvidenceSupport@$k$ (share of top-$k$ query-relevant claims with independent support) are utility/grounding
measures; a useful defense raises attack-arm reductions without false suppression or utility loss.
\subsection{The Defense Scoring Interface}
\label{sec:pf-score}
The defense consumes packet-relative claim labels. Let
$n_Q(c)=|G_{\mathrm{QR}}(c)|$ and
$n_Q^{\max}(q)=\max_{c\in C_q}n_Q(c)$ for the current rerank unit.
The normalized \emph{unsupported mass} is
\begin{equation}
 U(q,c)=\frac{\sum_{g\in G_{\mathrm{QR}}(c)}
 \mathbb{1}[\sigma_{E_q}(g)=\texttt{unsupported}]}
 {\max\{1,n_Q^{\max}(q)\}},
 \label{eq:umass}
\end{equation}
with $U=0$ for candidates without query-relevant claims.
Only \texttt{unsupported} enters the sum. Attested states, including
\codebreak{self\_claimed\_only}, never contribute; penalizing a listing's
own attested assertions would suppress honest self-description.
The \texttt{unverifiable} state is logged as abstention, not falsified. The denominator
is shared within a rerank unit but can vary between units; it is part of
the scoring rule, not absorbed into a single global $\lambda$.
Given the pointwise base score $s_0(q,c)$, the grounded score is
\begin{equation} s_G(q,c)=s_0(q,c)-\lambda\,U(q,c),\qquad \lambda\geq0, \label{eq:sg} \end{equation}
and candidates are re-sorted by descending $s_G$ under a fixed tie-break. $\lambda$ prices integrity against
utility; Section~\ref{sec:setup} specifies the development-split selection
rule and the fixed comparison grid. An identity precondition precedes any defense
number: at $\lambda{=}0$ the induced order must equal the base-score order
on every rerank unit. If $s_0$ comes from a scorer distinct from the phenomenon ranker, $r_0$ need not equal $r_m$; the orders
are reported separately, and the identity check concerns $s_G$ versus $s_0$ only.
This interface is post-hoc: it needs a numeric $s_0$ per candidate but does
not modify the base ranker. Section~\ref{sec:method} describes claim
extraction, matching, and the audit ledger.

\section{The GroundedGEO Framework}
\label{sec:method}

\subsection{Setup and Interface}

GroundedGEO is a claim-level, post-hoc reranking layer after a base
ranker. Inputs: query $q$, candidate set $C_q$, supplied evidence
packet $E_q$, and numeric base score $s_0(q,c)$.
Outputs: adjusted score $s_G(q,c)$, the induced order, and a
claim-level audit ledger (\S\ref{sec:method-match}), under the
closed-world label semantics of Section~\ref{sec:pf-evidence}.

\subsection{Pipeline Overview}

\begin{figure*}[t]
  \centering
  \includegraphics[width=\textwidth]{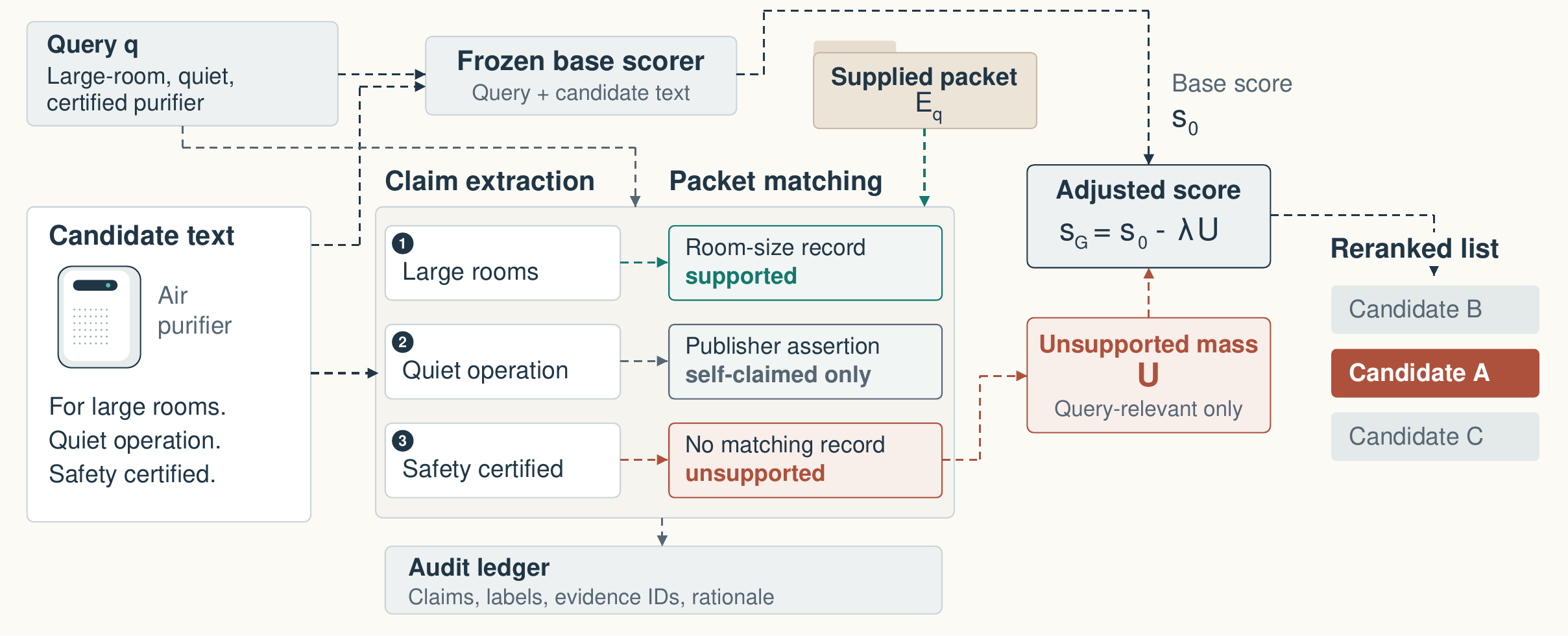}
  \caption{GroundedGEO's claim-level reranking interface. Query and candidate
  text feed the frozen base scorer; the supplied packet enters only claim
  matching. Only query-relevant \texttt{unsupported} claims contribute to
  $U$; all labels remain auditable. Three of the five label states are shown;
  \texttt{unverifiable} remains an abstention, not an unsupported verdict.
  The product, records, and output order are schematic. This diagram describes
  the interface, not a validated automatic defense.}
  \Description{A query and an air-purifier description feed a frozen base
  scorer that returns s-zero. A separate path extracts claims and matches
  them against a supplied evidence packet. Large-room suitability is supported
  by a room-size record, quiet operation is self-claimed only, and safety
  certification has no matching record and is unsupported. The unsupported
  query-relevant mass U combines with the base score as s-G equals s-zero
  minus lambda U, producing a reranked list. An audit ledger retains the
  claim labels, evidence identifiers, and rationales.}
  \label{fig:method-pipeline}
\end{figure*}

Figure~\ref{fig:method-pipeline} shows the two input paths.
The pipeline decomposes text into atomic, query-relevant claims
(\S\ref{sec:method-claims}); grounds each against the packet for a
five-way label and independence grade (\S\ref{sec:method-match});
aggregates unsupported mass for reranking (\S\ref{sec:method-score});
and retains the matching decisions for audit. Packet provenance makes
the evidence intervention replayable (\S\ref{sec:method-packets}).

\subsection{Atomic Claim Extraction}
\label{sec:method-claims}

An extractor emits claims $G(c)$ per candidate
(\S\ref{sec:pf-evidence}), each carrying a \emph{claim type} (six
categories: Specification, Use Case, Credential, Location\slash
Service Coverage, Case\slash Customer, Comparative), a \emph{query
relevance} grade $\rho_i\in\{\text{high},\text{medium},\text{low}\}$,
and an external-evidence flag. The \emph{high} tier forms
$G_{\mathrm{QR}}(c)$; other claims stay in the audit trace,
unpenalized. Benchmark candidates ship with claims and
\emph{construction} support labels as oracle five-way labels; the
automatic extractor instantiates the identical schema absent
provenance (the gated seam, \S\ref{sec:method-conditions}).

\subsection{Closed-World Evidence Matching}
\label{sec:method-match}

Claim $g_i$ is \emph{independently supported} when some packet atom
entails it and its source clears the independence threshold $\tau$
(\S\ref{sec:pf-evidence}):
\begin{equation}
 z_i = \max_{e_j\in E_q} \mathbb{1}[e_j\models g_i]\,\mathbb{1}[\operatorname{indep}(e_j)\geq\tau], \label{eq:support}
\end{equation}
A per-claim judge (not emitting $z_i$) sees the claim's
construction-linked atoms, else the whole packet up to a fixed cap;
absence of support in $E_q$ is itself the verdict. It returns a
five-way label (\S\ref{sec:pf-evidence}), an independence grade in
$\{0,\ldots,4\}$, and a short rationale---\texttt{unverifiable} is
retained, not coerced---and appends to the ledger the claim, label,
grade, shown evidence identifiers, model version, and parse status.

\subsection{Evidence Packets and Packet Twins}
\label{sec:method-packets}

Packet atoms retain their source type, independence tier, and identifier
in the matching ledger. The twin operation in Eq.~\ref{eq:twin} records
\codebreak{twin\_of}/\codebreak{twin\_profile} provenance so an audit can
replay the same candidate against the original and augmented packets.
Construction-tier tags distinguish injected attestations from retrieved
source records. Table~\ref{tab:packets} gives the benchmark's source layers.

\subsection{Grounded Scoring and Reranking}
\label{sec:method-score}

The implementation computes $U$ once per candidate using the maximum
query-relevant claim count in its rerank unit (Eq.~\ref{eq:umass}),
then sorts by descending $s_G$ (Eq.~\ref{eq:sg}); ties use ascending
candidate identifier. For nonempty claim sets, $U$ can equivalently be
computed as the unsupported share multiplied by $n_Q(c)/n_Q^{\max}(q)$.
This claim-count scaling is distinct from evidence-packet coverage.
Labels also yield a
supported-relevance share and a laundering-risk flag, diagnostics that
do not enter the score: laundering gets no bespoke patch; third-party
framing lacking an eligible third-party source enters $U$ like any
unattested claim.

\paragraph{Execution.} The base score is a pointwise 0--100 relevance
judgment per (query, candidate-text) pair; $\lambda$ is in score
points per unit unsupported mass. The pointwise seam is
primary---listwise rankers expose only an order (kept as a robustness
check). Handling is fail-closed: candidates lacking numeric $s_0$ are
dropped and counted, never imputed; the $\lambda{=}0$ identity
(rerank order equals base-score order) is enforced per unit by
refusing to emit rankings on failure.

\subsection{Reliability Conditions and Cost}
\label{sec:method-conditions}

Validity hinges on the matcher's reliability: the judge must pass a
frozen gate---macro-F1$\geq0.75$, Cohen's $\kappa\geq0.60$ on a
query-disjoint human-adjudicated holdout---before labels enter
confirmatory reranking; calibration uses the development split only.
Absent the gate, construction labels (oracle by provenance, not human
gold) bound what the mechanism \emph{could} deliver; automatic
reranking is diagnostic only. Cost is inference-only: no training, a judge
pass linear in claims, the supplied packet.

The interface assumes supplied evidence; open-Web discovery, entity
resolution, and universal fact checking are outside its scope.

\section{Experimental Setup}
\label{sec:setup}

\subsection{Research Questions}
\begin{description}
  \item[RQ1: Unsupported-specific promotion.] Do unsupported-rich
  variants gain rank over clean, unlike supported-rich and neutral arms?
  \item[RQ2: Evidence defense.] Does \method{} suppress
  unsupported promotion without falsely suppressing the other arms?
  \item[RQ3: Evidence sensitivity.] With text fixed, does changing the
  packet change the mechanism's output?
  \item[RQ4: Automatic evidence labeling (diagnostic).] Can a local
  language-model judge supply these labels, gated against human gold?
\end{description}

\subsection{Benchmark and Candidate Arms}
The benchmark comprises 50 English e-commerce queries derived from the
ESCI Shopping Queries corpus~\cite{reddy2022shopping}. Each query
realizes the five arms in Table~\ref{tab:variants}. Rich arms follow
three \emph{claim profiles} fixing the three
added claims' types (\textbf{S}: specification; \textbf{U}: use-case;
\textbf{SUC}: one per type), giving $1 + 4\times3 = 13$ variants per
query, for the 1{,}950 cases defined in Section~\ref{sec:pf-task}. Only
the target carries claims; the distractors carry none, so the penalty
$U(q,c)$ (Eq.~\ref{eq:umass}) can act on the target alone---the defense
lowers a target's rank rather than reordering competitors.

\paragraph{Evidence packets.} Every query has a versioned evidence
packet shared by all arms; atoms are assembled in three layers at
distinct independence tiers (Table~\ref{tab:packets}). One registry
case is contested (\texttt{q017}: GREENGUARD Gold claimed, contradicted
by independent sources).

\begin{table}[t]
  \centering\small
  \caption{Evidence-packet layers (independence: ordinal provenance).}\label{tab:packets}
  \begin{tabularx}{\columnwidth}{@{}lY@{}}
    \toprule
    Layer & Source type; independence; scale \\
    \midrule
    Listing & \codebreak{official\_public\_dataset\_record}; 1; 943 atoms, all 50 queries \\
    Buyer-review & \codebreak{platform\_user\_review} (Amazon Reviews 2023); 3; 13{,}265 reviews $\rightarrow$ 828 atoms, 42/50 ASINs \\
    Registry & \codebreak{certification\_registry}/\codebreak{registry\_corroboration}; 3--4; 13 atoms, 18 credential queries \\
    \bottomrule
  \end{tabularx}
\end{table}

\subsection{Label Sources and the Reliability Gate}
Claim labels come from four sources: construction-time labels, the
370-claim human gold, and two local judges. The API judge
(DeepSeek-Flash) is a holdout-only side check, not a fifth source
comparable to the local judges.

\paragraph{Construction labels.} The benchmark ships each claim with a
construction-time five-way label (oracle by provenance, not human gold).
These labels supply the reference reranking arm; their agreement with
human gold is evaluated alongside the automatic sources.

\paragraph{Human gold (370 claims).} A frozen 10-query holdout supplies
the human gold: 37 claims per query, 370 in total, each labeled
independently by two annotators (five-way evidence label; source
independence). Pre-adjudication $\kappa$: 0.833 (support), 0.816
(independence); 329 claims agreed outright, 41 adjudicated. Gold: 166
\texttt{unsupported}, 100 \texttt{supported}, 91
\codebreak{self\_claimed\_only}, 13 \codebreak{independent\_source\_supported},
0 \texttt{unverifiable}.

\paragraph{Local judges.} Two local judges label the same claims.
L1 is \codebreak{RedHatAI/Qwen3.6-35B-A3B-NVFP4} on the existing
\texttt{:8000} endpoint; L2 is \codebreak{unsloth/Qwen3.8-27B-NVFP4} on
\texttt{:8002} (NVFP4 quantisation, GPU utilisation $0.35$,
\codebreak{-{}-max-num-seqs 32}, \codebreak{enable\_thinking=false}). Each
labeled all 4{,}250 claims; parsing completeness and agreement with gold
are reported in Section~\ref{sec:res-matcher}.

\paragraph{API judge (holdout-only).} The API judge, DeepSeek-Flash,
was run only on the 10-query holdout ($850$ rows). Because it never
ran on the development split, its
labels cannot be treated as the same experiment as the local judges and
serve the gate check only. The same model also supplies the pointwise
base score $s_0(q,c)$ (Eq.~\ref{eq:sg}).

\paragraph{Reliability evaluation.} We apply the predeclared gate in
Section~\ref{sec:method-conditions} on the holdout. Alongside five-way
agreement, we report precision and recall of the binary penalty decision,
false penalties on attested claims, and both attested-only and operational
escape rates (Table~\ref{tab:label-source}).

\paragraph{Label-noise sweep.} To read the gate against endpoint needs
rather than accuracy alone, we resample the penalty-relevant binary
(\codebreak{unsupported} vs.\ attested) on the construction labels to
target precision $P$ and recall $R$, deterministically, and re-run the
reranker at $\lambda{=}40$ over all 1{,}950 cases. Precision targets are
$0.30$, $0.70$, $0.90$, and $1.00$; recall targets are $0.2$, $0.4$,
$0.6$, $0.8$, and $1.0$. The measured sources' binary precision and
recall provide reference points for interpreting the sweep.

\subsection{Probes: Packet Twins and Thinned Packets}
\paragraph{Packet twins.} 350 flipped packets ($50 \times 7$ claim
profiles, 1{,}050 injected claims) instantiate Eq.~\ref{eq:twin};
the S, U, and SUC profiles enter the core ranking cases. For RQ3 we
compare the original and twin packets at fixed text and $\lambda$.

\paragraph{Thinned packets.} The mirror probe removes the attesting
record of otherwise honest claims, so each thinned claim becomes
\codebreak{unsupported} under the closed-world rule and the penalty
appears. The dose axis is the benchmark-wide share of attested claims
whose attesting record is absent---$0/25/50/75/100\%$, i.e.\
$0/263/525/788/1050$ claims thinned; the listing layer is retained
throughout. Per-candidate fractions are degenerate (each candidate
carries three attested claims), so the share is pooled across the
benchmark. Thinned packets carry \codebreak{thinned\_of}/\codebreak{thinned\_level}
provenance, mirroring the twins, and all levels are reranked at
$\lambda{=}40$ over all 1{,}950 cases.

\subsection{Metrics and Their Link to the Estimand}
\paragraph{Phenomenon.} The phenomenon estimand is query-level NRG
(Eq.~\ref{eq:nrg}; seeds averaged within query, $n{=}50$).
\paragraph{Defense.} The main defense tables report each arm's raw
\emph{top-3 rate}, the mean of $S_3^d$ in Section~\ref{sec:pf-estimands}.
Attack suppression is its drop for unsupported-rich and
evidence-laundering (Success reduction); FSR is the drop for supported-rich,
neutral, and clean. These rates are distinct from the clean-referenced
promotion event $P_3^d$ and do not by themselves establish the frozen
Promotion reduction endpoint. Oracle reranking uses
$\lambda\in\{0,10,20,40\}$ over all 1{,}950 cases, subject to the
identity check in Section~\ref{sec:pf-score}. EDG remains descriptive.
\paragraph{Comparators.} A per-token length penalty (0.05, 0.10, 0.15)
on the same $s_0$ is the text-only comparator; for RQ3, candidates are
reranked under each twin packet at $\lambda{=}40$.
\paragraph{Inference.} Wilcoxon signed-rank tests on paired query-level
contrasts, Holm-corrected within each claim profile. RQ1's primary
contrasts are unsupported-rich versus clean, supported-rich, and
neutral, with the remaining arms versus clean as specificity controls.

\subsection{Split, Tuning Rule, and Implementation}
\paragraph{Frozen split.} The 50 queries are partitioned into a
40-query development split and a 10-query holdout, query-disjoint.
The split is frozen before evaluation: the development split is used
to evaluate the $\lambda$ selection rule and for calibration; holdout
labels are not used to select $\lambda$.
All label-source validation, the reliability gate, and the 370-claim
human gold live on the holdout-10.
\paragraph{Pre-registered tuning rule.} The rule below is
pre-registered for this revision; it was not part of earlier drafts.
Rule text: among $\lambda$ with $\text{FSR}=0$, select the smallest
$\lambda$ that attains $95\%$ of the maximum suppression (the knee); if
no $\lambda$ attains $\text{FSR}=0$, select the $\lambda$ with the
smallest FSR and flag it as a fallback. FSR is the binding constraint in
the rule. Reported operating points are
$\lambda\in\{0,10,20,40\}$ (the main comparisons use $\lambda{=}40$ for
comparability with the oracle curve); the rule was additionally
evaluated at $\lambda\in\{80,160\}$. We report the swept points and
flag grid-edge selections; the main $\lambda=40$ comparisons are not
presented as evaluations of a development-selected optimum.
\paragraph{Implementation and reproducibility.} Queries, candidates,
packets, claim ledgers, and order manifests are versioned and
hash-pinned; model identifiers, prompts, and parser schemas are frozen
pre-evaluation. The auxiliary rankers that produce the phenomenon
measurements are Qwen2.5-7B (local), GLM-5.3-Flash, and MiMo-v2.5
(API)~\cite{pradeep2023rankvicuna,zhang2023rankwithoutgpt}, all
receiving identical candidate text. A result manifest binds each table
to input hashes, raw logs, and analysis code.

\section{Results}
\label{sec:results}
\sloppy
Unless a subset is named, results use the frozen 50-query benchmark
(1{,}950 cases) and the protocol of Section~\ref{sec:setup}. Phenomenon
results use the three listwise rankers; all defense comparisons share
the pointwise base scorer and report raw top-3 rates and their drops.
Automatic-label, coverage, and label-noise arms are diagnostic.

\subsection{Unsupported-Specific Rank Promotion (RQ1)}
\label{sec:res-promotion}
\begin{figure*}[t]
  \centering
  \includegraphics[width=\textwidth]{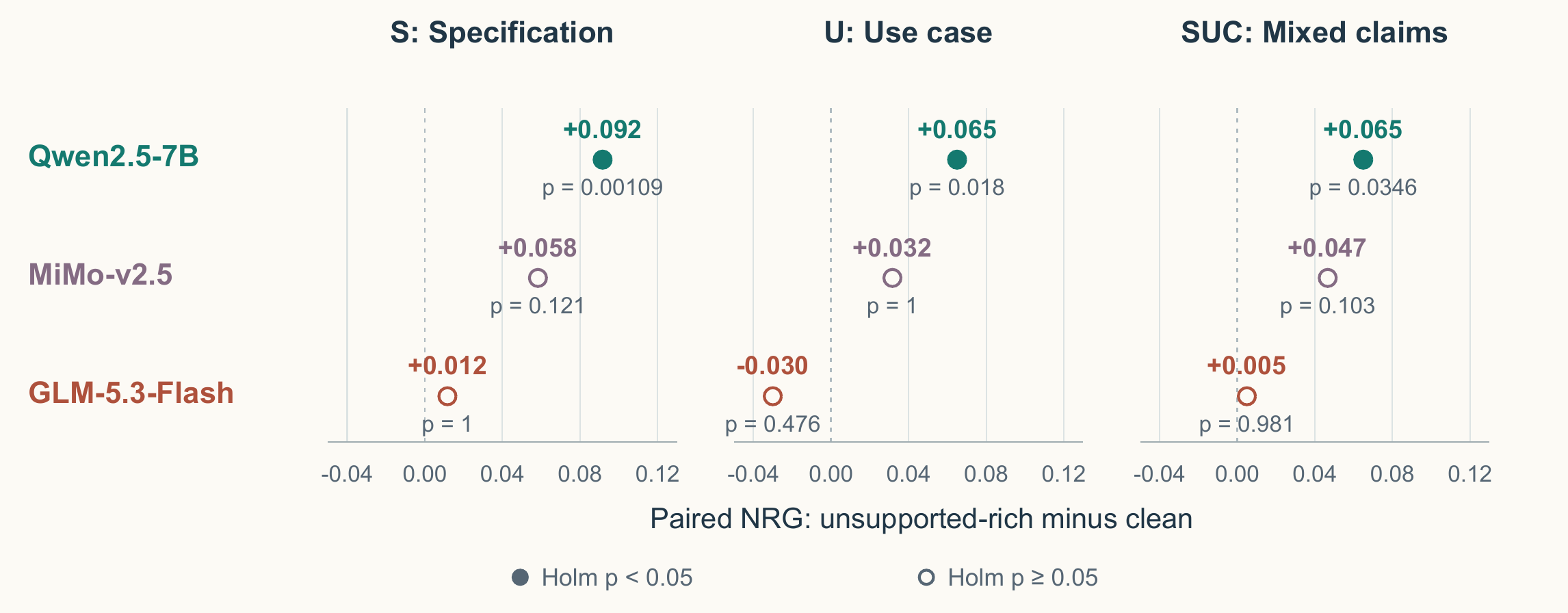}
  \caption{Model-specific promotion of unsupported-rich over clean, using
  the frozen query-level paired statistics ($n=50$, seeds averaged within query).
  Columns are specification (S), use-case (U), and mixed (SUC) profiles.
  Points show mean paired NRG; filled points have Holm-adjusted $p<0.05$.
  All displayed $p$ values are within-model, within-profile tests, not tests
  of differences between models. Other comparators, including MiMo's
  significant S-profile contrast against neutral-matched, are in
  Table~\ref{tab:promotion}.}
  \Description{Three panels plot paired normalized rank gain against clean.
  Qwen2.5-7B has gains 0.092, 0.065, and 0.065, all significant after Holm
  correction. MiMo-v2.5 has 0.058, 0.032, and 0.047, none significant for
  this comparator. GLM-5.3-Flash has 0.012, minus 0.030, and 0.005, also
  nonsignificant. Positive and negative values are drawn on identical axes.}
  \label{fig:model-susceptibility}
\end{figure*}

Qwen2.5-7B promotes unsupported-rich over clean on all three claim
profiles (paired NRG $+0.065$ to $+0.092$, Holm-adjusted $p$ between
$0.035$ and $0.0011$; Table~\ref{tab:promotion}); at S it even beats
supported-rich ($+0.103$, $p{=}0.0010$)---fabricated richness outranks
honest richness. The gain is unsupported-specific: supported-rich,
neutral-matched, and laundering arms never separate from clean.
Figure~\ref{fig:model-susceptibility} isolates the unsupported-rich versus
clean contrast across rankers and profiles.
\begin{table*}[t]
  \centering\small
  \caption{Paired NRG contrasts (Eq.~\ref{eq:nrg}); Holm-adjusted $p$ in
  parentheses, \textbf{bold} $p<0.05$; controls n.s.\ (min $0.064$).}
  \label{tab:promotion}
  \begin{tabular}{@{}lllccc@{}}
    \toprule
    Ranker & Arm & Comparator & S & U & SUC \\
    \midrule
    \multirow{6}{*}{Qwen2.5-7B}
      & unsupported-rich & clean & \textbf{+0.092}~(0.0011) & \textbf{+0.065}~(0.018) & \textbf{+0.065}~(0.035) \\
      & unsupported-rich & supported-rich & \textbf{+0.103}~(0.0010) & $+0.055$~(0.063) & $+0.045$~(0.059) \\
      & unsupported-rich & neutral-matched & \textbf{+0.100}~(0.0011) & \textbf{+0.063}~(0.049) & $+0.058$~(0.069) \\
      \cmidrule(l){2-6}
      & supported-rich & clean & $-0.012$ & $+0.010$ & $+0.020$ \\
      & neutral-matched & clean & $-0.008$ & $+0.002$ & $+0.007$ \\
      & evidence-laundering & clean & $+0.042$ & $+0.023$ & $+0.018$ \\
    \midrule
    \multirow{2}{*}{MiMo-v2.5}
      & unsupported-rich & clean & $+0.058$~(0.121) & $+0.032$~(1.0) & $+0.047$~(0.103) \\
      & unsupported-rich & neutral-matched & \textbf{+0.068}~(0.037) & -- & -- \\
    \midrule
    GLM-5.3-Flash & \multicolumn{5}{p{0.58\textwidth}@{}}{no significant contrast in any profile (21 tests after Holm, incl.\ negative directions; e.g., SUC neutral-matched vs.\ clean $-0.030$, $p{=}0.41$)} \\
    \bottomrule
  \end{tabular}
\end{table*}
MiMo-v2.5 has one significant corrected contrast (S-profile
unsupported-rich versus neutral-matched); GLM-5.3-Flash has none.
These within-model tests establish where promotion is detected, not a
formal difference between models or equivalence on nonsignificant arms.
On Qwen, the significant matched-rich contrasts are stronger evidence
against a length-only explanation than the comparisons with clean:
the rich arms share a length constraint, whereas clean does not. The
claim is therefore a susceptibility demonstrated on particular rankers,
not a universal consequence of adding detail.

\subsection{Text-Only Proxies Cannot Separate Honest from Fabricated Richness (RQ2)}
\label{sec:res-textproxy}
The per-token length penalty on the same base scorer cannot reproduce
evidence-conditioned suppression. From $0.05$ to $0.10$ per token nothing
moves (supported/neutral $0.75$, unsupported $0.65$, laundering $0.61$); at
$0.15$ the attack arms drop only to $0.63$ and $0.59$ while the honest rich
arms fall to $0.73$ (Table~\ref{tab:length-baseline}).
\begin{table}[t]
  \centering\small
  \caption{Per-token length penalty on the same base scorer $s_0$, per-arm
  top-3 rate; the text-only comparator to
  Table~\ref{tab:oracle-rerank}.}
  \label{tab:length-baseline}
  \begin{tabular}{@{}lcccc@{}}
    \toprule
    Penalty (per token) & supported & unsupported & laundering & neutral \\
    \midrule
    $0.05$ & 0.75 & 0.65 & 0.61 & 0.75 \\
    $0.10$ & 0.75 & 0.65 & 0.61 & 0.75 \\
    $0.15$ & \textbf{0.73} & 0.63 & 0.59 & \textbf{0.73} \\
    \bottomrule
  \end{tabular}
\end{table}
The weak settings leave the ranking unchanged; the stronger setting
reduces both attack and protected rich-arm rates by about the same amount.
Thus increasing the proxy penalty does not supply the missing distinction
at the tested strengths. This comparison is specific to the length proxy
on matched arms, not evidence that every text-only defense is ineffective
on every attack. The fixed-text twin probe below isolates the broader
information limitation.

\subsection{Oracle Evidence-Constrained Reranking (RQ2)}
\label{sec:res-oracle}
\begin{figure*}[t]
  \centering
  \includegraphics[width=\textwidth]{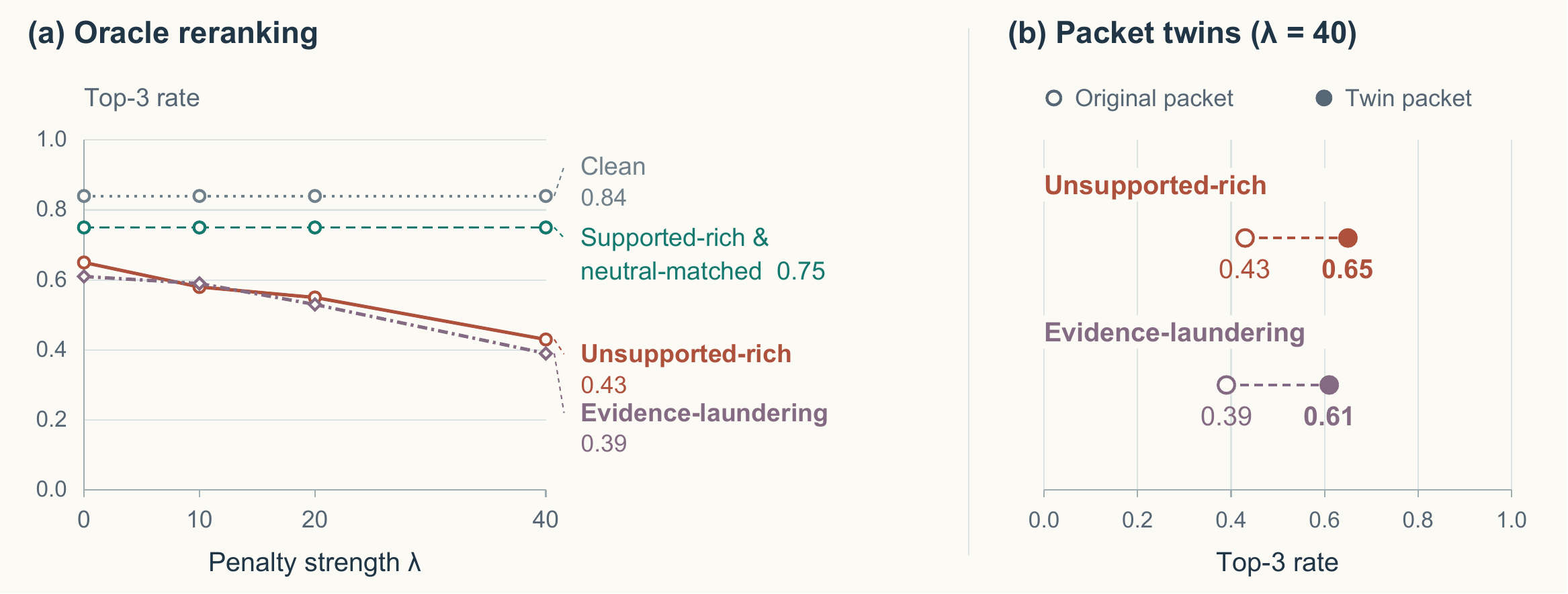}
  \caption{Oracle reranking and packet twins, using the rounded rates in
  Table~\ref{tab:oracle-rerank} (1,950 cases).
  (a) Lines connect tested penalty strengths; the supported-rich and
  neutral-matched series overlap.
  (b) Original and synthetic twin packets at fixed $\lambda=40$ and text.
  Hollow and filled points denote original and twin packets, respectively.
  These construction-label results bound the mechanism, not validated
  automatic-defense performance.}
  \Description{Two panels show oracle top-3 rates. The left plots penalty
  strengths zero, 10, 20, and 40 on a numeric axis. Clean remains 0.84;
  supported-rich and neutral-matched both remain 0.75. Unsupported-rich
  follows 0.65, 0.58, 0.55, and 0.43; evidence-laundering follows 0.61,
  0.59, 0.53, and 0.39. The right compares original and twin packets at
  penalty strength 40: unsupported-rich recovers from 0.43 to 0.65, and
  evidence-laundering from 0.39 to 0.61.}
  \label{fig:oracle-results}
\end{figure*}

Under oracle labels, \method{}'s penalty suppresses both attack
arms: as $\lambda$ grows $0\to40$, unsupported-rich falls monotonically
$0.65\to0.43$ and laundering $0.61\to0.39$, while clean ($0.84$),
supported-rich ($0.75$), and neutral ($0.75$) stay flat at every
$\lambda$---FSR zero (Figure~\ref{fig:oracle-results}a;
Table~\ref{tab:oracle-rerank}).
\begin{table}[t]
  \centering\small
  \caption{Oracle \method{} reranking ($s_G{=}s_0-\lambda U$), 1{,}950
  cases, per-arm top-3 rate; twins flip only the packet.}
  \label{tab:oracle-rerank}
  \begin{tabular}{@{}lccccc@{}}
    \toprule
    Arm & $\lambda{=}0$ & $\lambda{=}10$ & $\lambda{=}20$ & $\lambda{=}40$ & twin $\lambda{=}40$ \\
    \midrule
    clean               & 0.84 & 0.84 & 0.84 & 0.84 & 0.84 \\
    supported-rich      & 0.75 & 0.75 & 0.75 & 0.75 & 0.75 \\
    neutral-matched     & 0.75 & 0.75 & 0.75 & 0.75 & 0.75 \\
    unsupported-rich    & 0.65 & 0.58 & 0.55 & \textbf{0.43} & \textbf{0.65} \\
    evidence-laundering & 0.61 & 0.59 & 0.53 & \textbf{0.39} & \textbf{0.61} \\
    \bottomrule
  \end{tabular}
\end{table}
The $\lambda=0$ identity check passed on all 1{,}950 cases, excluding
base-order mismatch as an explanation. Under construction labels the
protected arms carry no unsupported mass, so their zero FSR follows from
the rule rather than from threshold calibration; honest self-descriptions
are not penalized. The length comparator lacks this selectivity.
This establishes a mechanism-level reference under the supplied labels,
not automatic-label reliability or protection when evidence is missing;
the following experiments test those dependencies separately.

\subsection{Packet Twins: An Existence Proof for Evidence Worlds (RQ3)}
\label{sec:res-twins}
Flipping only the evidence world---twin packets whose injected atoms
attest the previously unsupported claims---restores the attack arms
exactly to $\lambda{=}0$ rates (unsupported-rich $0.43\to0.65$,
laundering $0.39\to0.61$; Figure~\ref{fig:oracle-results}b;
Table~\ref{tab:oracle-rerank} last column; $\lambda{=}40$).
The injected attestations remove the penalized mass without changing
$s_0$. Their construction-tier tags do not confer authoritative source
status, so the contrast isolates attestation availability rather than
source prestige. A function of $(q,c)$ alone cannot reproduce this
fixed-text difference. The result is an existence probe for the evidence
channel, not a test of real-world retrieval or certificate authentication.

\subsection{The Automation Gap: Claim--Evidence Matching Is the Bottleneck (RQ4)}
\label{sec:res-matcher}
All three automatic judges fail the preregistered reliability gate against
human gold; the best local judge reaches macro-F1 $0.374$ and
$\kappa$ $0.442$, well under the gate's $0.75/0.60$. Because the labels
that carry the oracle result are construction-time, this is a separate
test of their automatic reproduction. Table~\ref{tab:label-source}
compares all sources against the 370-claim gold;
Table~\ref{tab:matcher-gate} gives complementary agreement and binary
diagnostics. Both local judges parsed all 4{,}250 claims successfully.
\begin{table}[t]
  \centering\small
  \caption{Complementary diagnostics on the 370-claim holdout:
  pre-adjudication human agreement and API-judge binary performance.
  Five-way and penalty-decision metrics appear in
  Table~\ref{tab:label-source}.}
  \label{tab:matcher-gate}
  \begin{tabular}{@{}lc@{}}
    \toprule
    Quantity & Value \\
    \midrule
    Annotator agreement $\kappa$ (support / independence) & 0.833 / 0.816 \\
    Judge binary accuracy & 0.738 \\
    Attested recall & 0.975 \\
    \bottomrule
  \end{tabular}
\end{table}
\begin{table*}[t]
  \centering\small
  \caption{Label sources against the 370-claim human gold (holdout-10),
  $\lambda{=}40$ inputs. Gate: macro-F1 $\geq0.75$ \emph{and} Cohen's
  $\kappa\geq0.60$; all three automatic judges fail. ``P/R$_{\text{pen}}$''
  is precision/recall of the penalty (unsupported) decision; FSR is
  attested claims penalized; the two escape columns are the attested-only
  and operational-definition escape rates. The API judge covers only the
  850 holdout rows and is not the same experiment as the local judges.}
  \label{tab:label-source}
  \begin{tabular}{@{}lcccccccc@{}}
    \toprule
    Source & 5-way acc & macro-F1 & $\kappa$ & P$_{\text{pen}}$ & R$_{\text{pen}}$ & FSR & esc.\ (attested) & esc.\ (op.) \\
    \midrule
    construction (oracle row) & 0.908 & 0.692 & 0.859 & 0.894 & 0.970 & 0.093 & 0.030 & 0.030 \\
    DeepSeek-Flash (API)      & 0.419 & 0.246 & 0.213 & 0.957 & 0.271 & 0.010 & 0.554 & 0.729 \\
    L1 (\texttt{Qwen3.6-35B-A3B-NVFP4}) & 0.522 & 0.276 & 0.301 & 0.919 & 0.548 & 0.039 & 0.446 & 0.452 \\
    L2 (\texttt{Qwen3.8-27B-NVFP4})     & 0.630 & \textbf{0.374} & \textbf{0.442} & 0.944 & \textbf{0.819} & 0.039 & 0.181 & 0.181 \\
    \bottomrule
  \end{tabular}
\end{table*}
\paragraph{Error direction.}
The error direction is uniformly credulous: precision on the penalty
decision is high (L2 $0.944$) while recall is the binding weakness
(L2 $0.819$ best, API $0.271$ worst), so fabricated claims are passed
rather than honest ones suppressed. The two escape columns must be read
separately---the API judge abstains heavily ($0.554\to0.729$ under the
operational definition) whereas L2 does not ($0.181$/$0.181$). The
strongest alternative---that the gate simply reflects noisy gold---is
weakened by the construction row, which scores $\kappa$ $0.859$ with
$3.0\%$ escape and $9.3\%$ false suppression against the same gold: the
gold can be matched closely, and the automatic judges fail directionally,
not randomly.
These are point estimates on one 10-query holdout, not interval estimates.
The API judge has no development-split run and cannot supply the full-set
end-to-end comparison below. The gate remains unmet; we next examine
ranking behavior diagnostically without relaxing that admission rule.

\subsection{End-to-End Arms: What the Automatic Labels Actually Do (RQ2, diagnostic)}
\label{sec:res-e2e}
At $\lambda{=}40$ on the same pointwise base scorer, oracle labels suppress
the attack arms with zero false suppression; both local judges also leave
the protected arms untouched ($\text{FSR}=0$) while suppressing
substantially, with L2 reaching $79\%$ of oracle unsupported-rich
suppression and $100\%$ of oracle laundering suppression on the full
1{,}950 cases, and $85$--$100\%$ on the shared 10-query subset
(Tables~\ref{tab:e2e-arms} and~\ref{tab:e2e-samecases}). The human-gold arm
is the one adverse-measured cell: its collateral is positive under the
current gold version, and its interpretation is pending.
\begin{table}[t]
  \centering\small
  \caption{End-to-end arms over all 1{,}950 cases, per-arm top-3 rate. The
  $\lambda{=}0$ column is shared by all sources (identical base order); the
  oracle, L1, and L2 columns are at $\lambda{=}40$. The oracle column
  reproduces the $\lambda{=}40$ rows of Table~\ref{tab:oracle-rerank}.}
  \label{tab:e2e-arms}
  \begin{tabular}{@{}lcccc@{}}
    \toprule
    Arm & $\lambda{=}0$ & oracle & L1 & L2 \\
    \midrule
    clean               & 0.840 & 0.840 & 0.840 & 0.840 \\
    supported-rich      & 0.753 & 0.753 & 0.753 & 0.753 \\
    neutral-matched     & 0.753 & 0.753 & 0.753 & 0.753 \\
    unsupported-rich    & 0.647 & \textbf{0.427} & \textbf{0.500} & \textbf{0.473} \\
    evidence-laundering & 0.613 & \textbf{0.387} & \textbf{0.420} & \textbf{0.387} \\
    \bottomrule
  \end{tabular}
\end{table}
\begin{table}[t]
  \centering\small
  \caption{Same cases (holdout-10, 390 cases), all four label sources at
  $\lambda{=}40$; the human-gold column is the only adverse-measured cell
  and its interpretation is pending (see text and
  Section~\ref{sec:limitations}).}
  \label{tab:e2e-samecases}
  \begin{tabular}{@{}lcccc@{}}
    \toprule
    Arm & oracle & human gold & L1 & L2 \\
    \midrule
    clean               & 0.700 & \textbf{0.400} & 0.700 & 0.700 \\
    supported-rich      & 0.733 & 0.733 & 0.733 & 0.733 \\
    neutral-matched     & 0.700 & 0.700 & 0.700 & 0.700 \\
    unsupported-rich    & 0.233 & 0.500 & 0.400 & 0.300 \\
    evidence-laundering & 0.300 & 0.400 & 0.400 & 0.300 \\
    \bottomrule
  \end{tabular}
\end{table}
\paragraph{Suppression and collateral summary (same cases).}
Attack suppression ($\lambda{=}0-\lambda{=}40$, unsupported / laundering)
and maximum collateral (FSR on protected arms): oracle $+0.433/+0.267$,
FSR $0.000$; human gold $+0.167/+0.167$, FSR $+0.300$; L1 $+0.267/+0.167$,
FSR $0.000$; L2 $+0.367/+0.267$, FSR $0.000$.
\paragraph{Comparison scope.}
L1 and L2 each cover all 50 queries; the human-gold arm covers only the
10-query holdout, so it is compared on the subset where all four sources
exist; the two local judges come from the same model family (Section~\ref{sec:setup}), so
agreement between them is not independent evidence.
\paragraph{Pending human-gold interpretation.}
The human-gold arm differs from the construction and local-judge arms:
the entire measured collateral
traces to five clean base-identity claims that the humans judged
\texttt{unsupported} (the expected construction label is
\texttt{self\_claimed\_only}), and part of the weaker suppression reflects
about $10\%$ of attack-arm claims that the humans judged attested, a
standards divergence distinct from the base-identity case. Whether the
collateral reflects gold noise or a definitional split is not adjudicated
here: these are the numbers \emph{as measured under this gold version}, and
their interpretation is marked pending (Section~\ref{sec:limitations}).
The local judges' favorable ranking effects do not reverse their gate
failure. Section~\ref{sec:discussion} considers why agreement on the
audit ledger and changes in ranking need not track one another.

\subsection{Coverage Dose--Response: The Cost of Unsupported Truth (diagnostic)}
\label{sec:res-coverage}
\begin{figure}[t]
  \centering
  \includegraphics[width=\columnwidth]{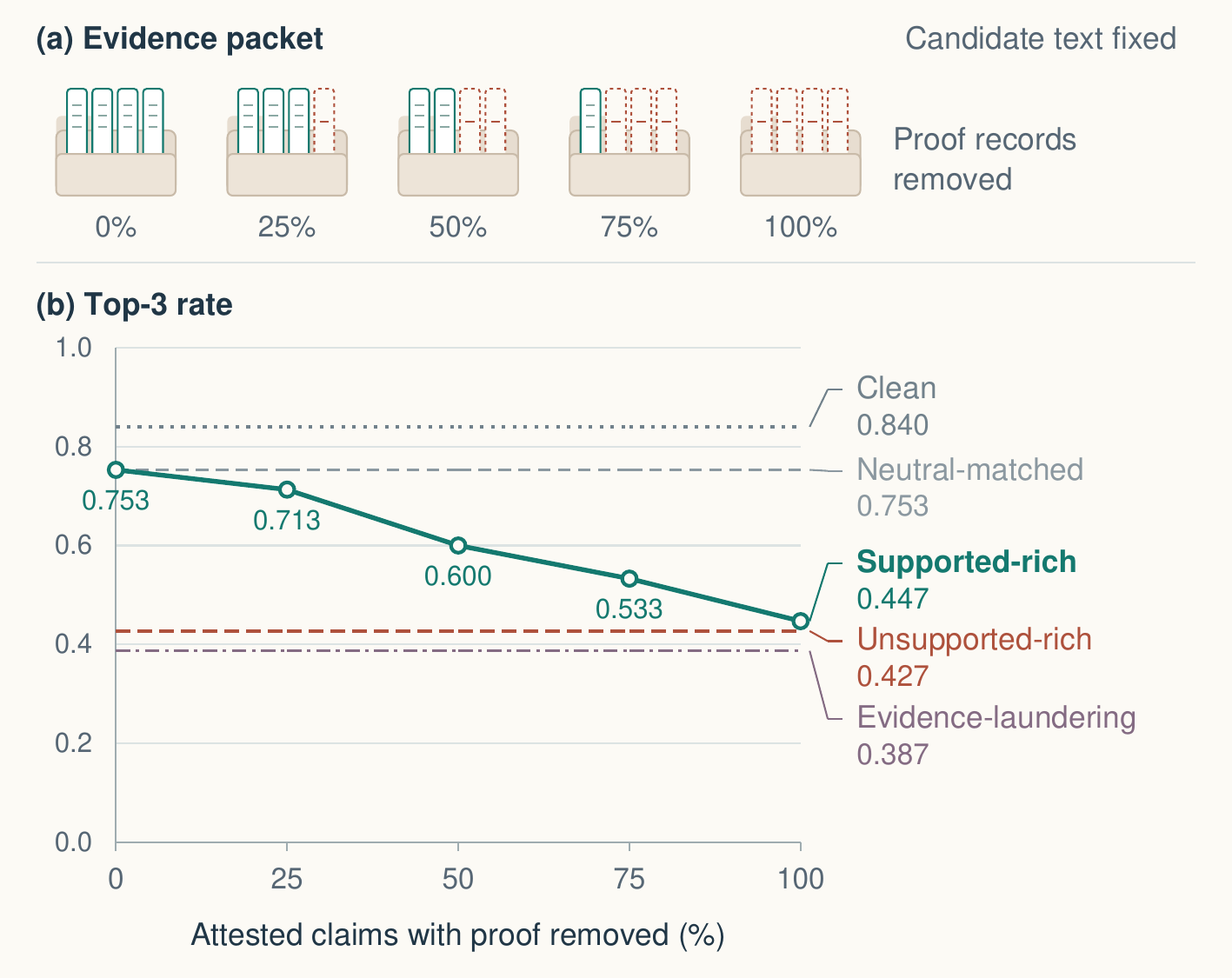}
  \caption{Evidence coverage dose--response ($\lambda=40$, 1,950 cases).
  (a) Packet thinning, with schematic proportions rather than record counts.
  (b) Measured rates from Table~\ref{tab:coverage}; lines connect tested doses.
  Text and the base scorer remain fixed. This controlled probe does not estimate
  deployed-system performance.}
  \Description{Five folders depict zero, 25, 50, 75, and 100 percent
  removal of proof records. Below, a line chart plots top-3 rate against
  the fraction of attested claims whose proof is removed. Supported-rich
  falls from 0.753 through 0.713, 0.600, and 0.533 to 0.447.
  Clean stays at 0.840, neutral-matched at 0.753, unsupported-rich at
  0.427, and evidence-laundering at 0.387.}
  \label{fig:coverage-dose}
\end{figure}

Thinning the packet---removing the proof records of an increasing fraction
of attested claims---monotonically depresses the honest supported-rich arm
while leaving clean, neutral, and both attack arms flat; at $100\%$
thinning an honest candidate ($0.447$) approaches a
fabricated one ($0.427$) at the endpoint
(Figure~\ref{fig:coverage-dose}; Table~\ref{tab:coverage}).
\begin{table}[t]
  \centering\footnotesize\setlength{\tabcolsep}{4pt}
  \caption{Coverage dose--response on thinned packets ($\lambda{=}40$,
  1{,}950 cases). Axis: fraction of attested claims whose proof record is
  not in the packet ($0/25/50/75/100\%\to 0/263/525/788/1050$ claims).
  FSR is the supported-rich top-3 drop relative to the unthinned packet.}
  \label{tab:coverage}
  \begin{tabular}{@{}lcccccc@{}}
    \toprule
    Thinned & supported-rich & FSR & neutral & clean & unsupported & laundering \\
    \midrule
    $0\%$    & 0.753 & 0.000 & 0.753 & 0.840 & 0.427 & 0.387 \\
    $25\%$   & 0.713 & $+0.040$ & 0.753 & 0.840 & 0.427 & 0.387 \\
    $50\%$   & 0.600 & $+0.153$ & 0.753 & 0.840 & 0.427 & 0.387 \\
    $75\%$   & 0.533 & $+0.220$ & 0.753 & 0.840 & 0.427 & 0.387 \\
    $100\%$  & \textbf{0.447} & \textbf{$+0.307$} & 0.753 & 0.840 & 0.427 & 0.387 \\
    \bottomrule
  \end{tabular}
\end{table}
The intervention changes evidence availability, not candidate length or
format: withdrawing an attestation changes its claim's label and hence
$U$. The other arms' labels are unchanged. At full thinning, supported-rich
incurs $+0.307$ FSR despite unchanged text. Because the same queries are
re-evaluated at each dose, the contrast avoids the query-difficulty
confound in comparing naturally covered and uncovered strata. It measures
the cost of engineered evidence loss, not its prevalence in deployment.

\subsection{Label-Noise Sweep: Precision Buys Safety, Recall Buys Suppression (diagnostic)}
\label{sec:res-noise}
\begin{figure*}[t]
  \centering
  \includegraphics[width=\textwidth]{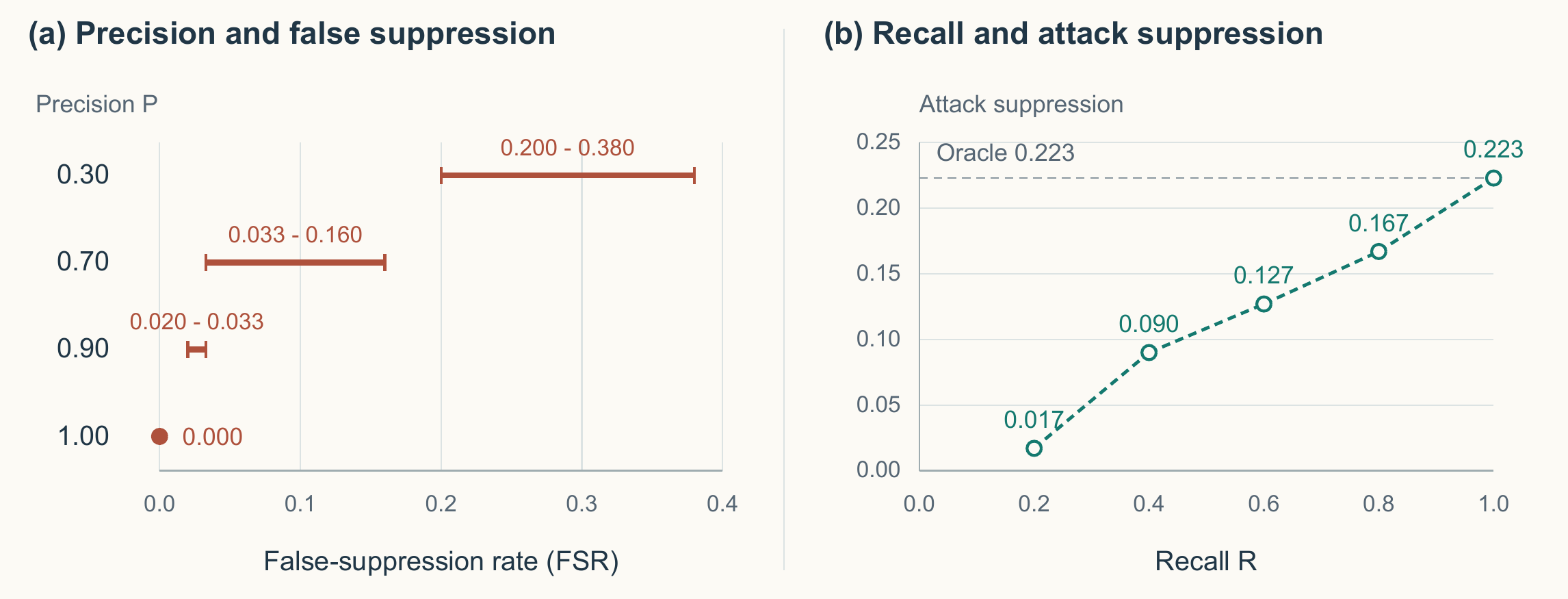}
  \caption{Label-noise diagnostics at $\lambda=40$ on 1,950 cases
  (Table~\ref{tab:noise}). (a) Horizontal segments show the reported FSR
  range across swept recall settings at each precision; these are simulation
  ranges, not confidence intervals, and no midpoint estimate is implied.
  (b) Attack suppression versus recall; dashed lines connect tested settings,
  with the oracle reference at 0.223. This deterministic relabeling simulation
  does not measure repeated-judge variability or validate an automatic judge.}
  \Description{At precision 0.30 the false-suppression range is 0.200 to
  0.380; at 0.70 it is 0.033 to 0.160; at 0.90 it is 0.020 to 0.033;
  at 1.00 it is zero. At recalls 0.2, 0.4, 0.6, 0.8, and 1.0,
  attack suppression is 0.017, 0.090, 0.127, 0.167, and 0.223.}
  \label{fig:label-noise}
\end{figure*}

Resampling the binary unsupported-versus-attested decision shows the two
error types act on different arms: label \emph{precision} controls
collateral on protected arms, and label \emph{recall} controls attack
suppression, whose ceiling is the oracle level $+0.223$
(Table~\ref{tab:noise}). The four measured label sources sit inside the
grid.
Figure~\ref{fig:label-noise} separates the two diagnostic axes.
\begin{table}[t]
  \centering\footnotesize
  \caption{Deterministic label-noise sweep at $\lambda{=}40$ (1{,}950
  cases): global re-sampling of the binary unsupported-versus-attested
  decision on the construction labels. Left and right column pairs
  summarize separate axes, not matched $(P,R)$ settings per row.
  Measured-source precision and recall are in Table~\ref{tab:label-source}.}
  \label{tab:noise}
  \setlength{\tabcolsep}{3pt}
  \begin{tabular}{@{}lclc@{}}
    \toprule
    Precision $P$ & FSR span & Recall $R$ & Suppression \\
    \midrule
    $0.30$ & $0.20$--$0.38$   & $0.2$ & $+0.017$ \\
    $0.70$ & $0.033$--$0.16$  & $0.4$ & $+0.090$ \\
    $0.90$ & $0.020$--$0.033$ & $0.6$ & $+0.127$ \\
    $1.00$ & $0.000$          & $0.8$ & $+0.167$ \\
           &                  & $1.0$ & $+0.223$ \\
    \bottomrule
  \end{tabular}
\end{table}
Relative to the oracle $(P,R)=(1,1)$, false unsupported labels add mass
to protected candidates, whereas missed unsupported labels remove mass
from attack candidates. This directional structure explains why aggregate
label accuracy alone is insufficient for predicting endpoint behavior.
The sweep is deterministic relabeling, not repeated sampling from a judge;
its ranges describe the tested settings and carry no estimate of judge
variability. It motivates separate safety and suppression requirements
without replacing the preregistered agreement gate.

\subsection{The \texorpdfstring{$\lambda$}{lambda} Tuning Curve Is Weakly Identified}
\label{sec:res-tuning}
Under the preregistered knee rule (Section~\ref{sec:setup}), the
operating point is weakly identified: the FSR constraint is nearly empty
on the dev-40 split (Table~\ref{tab:tuning}). The oracle
and L2 curves are selected at the grid edge ($\lambda{=}160$, flagged
\texttt{grid\_edge}); L1 selects $\lambda{=}40$.
\begin{table}[t]
  \centering\small
  \caption{$\lambda$ tuning on the dev-40 split under the preregistered knee
  rule. Dashes are grid points not swept per source; \texttt{grid\_edge}
  flags a selection at the largest tested strength. Main comparisons
  retain $\lambda=40$, not the selected grid-edge value.}
  \label{tab:tuning}
  \begin{tabular}{@{}lcccc@{}}
    \toprule
    Source & $\lambda{=}40$ supp. & $\lambda{=}80$ & $\lambda{=}160$ & selected \\
    \midrule
    oracle & 0.192 & 0.567 & 0.633 & 160 (\texttt{grid\_edge}, FSR $0$) \\
    L1     & --    & --    & --    & 40 (supp.\ 0.158, FSR $0$) \\
    L2     & --    & --    & --    & 160 (\texttt{grid\_edge}, FSR $0$) \\
    \bottomrule
  \end{tabular}
\end{table}
With almost no protected-arm error to trade against suppression, the rule
has little basis for selecting an interior operating point. A grid-edge
selection therefore does not identify a deployable optimum. We report the
swept points and selected settings, while retaining $\lambda=40$ as the
common comparison point; the main tables are not a tuned-and-frozen
holdout evaluation of that value.
\fussy

\section{Analysis and Discussion}
\label{sec:discussion}

The joint results distinguish three questions that a single ranking score
would conflate: whether a ranker rewards unsupported detail, whether a
defense can respond to evidence, and whether the evidence labels are
reliable enough to justify its decisions. A positive answer to the second
does not establish the third. This distinction explains both the useful
automatic-label ranking effects and the limits of interpreting them as a
validated defense.

\subsection{Mechanism}

\paragraph{Evidence access versus evidence use.}
The API judge's high attested recall coexists with substantial escape of
unsupported claims (Tables~\ref{tab:matcher-gate} and~\ref{tab:label-source}).
Unlike the text-only ranker, this judge receives evidence, so its errors
are not forced by the same information restriction. They are consistent
with a verifier accepting plausibility or apparent attestation in place of
packet entailment, but the aggregate errors do not establish that cause.
The practical implication is to audit the matching decisions themselves:
adding an evidence channel is necessary for the twin distinction, but
merely supplying a packet is not a reliability guarantee.

\paragraph{The scoring rule's normalisation sets where it bites.}
In Eq.~\ref{eq:umass}, the denominator is the largest query-relevant claim
count in the current rerank unit. Here only the target carries claims, so
for a nonempty target $n_Q^{\max}(q)=n_Q(t)$ and $U$ reduces to its
unsupported fraction. One disputed claim can therefore carry a large
penalty on a sparse description. This explains the score's sensitivity
to the clean-arm disagreements in Section~\ref{sec:res-e2e}, but does not
decide which labels are correct. If distractors also carried claims, the
denominator need not equal the target's count; this particular
rate interpretation would not generalize unchanged.

\paragraph{Coverage and matching require separate audits.}
The thinning probe intervenes upstream of the judge: even a matcher that
perfectly follows the closed-world rule must change a label when its sole
attestation disappears. Improving agreement with that rule therefore
cannot, by itself, protect uncovered truth. A deployment would need to
distinguish a failed match against adequate evidence from an inadequately
assembled packet, with retrieval or human review as possible responses to
the latter. These are deployment requirements suggested by the probe, not
capabilities validated by the current supplied-packet experiments.

\paragraph{``Attested'' is itself contested at the margin.}
Construction labels never assign \codebreak{independent\_source\_supported},
whereas the human annotators assign it to $13$ claims
(unsupported-rich~3, neutral~5, laundering~5), so agreement on independent
support is not automatic and the attested/unsupported boundary is a judgment
at the margin. Source eligibility and the binary decision to penalize
are consequently different annotation tasks; a ledger can disagree on
independence even when the ranking penalty is unchanged.

\subsection{Unresolved Alternatives}

\paragraph{The gate's form versus endpoint behaviour.}
The strongest reading against our automation gap is that it measures the
gate rather than the judge: L2's penalty decisions yield useful ranking
effects even though its five-way agreement is insufficient. The label-noise
sweep explains why a binary penalty can tolerate some errors among
non-penalized states while remaining sensitive to the direction of errors
across the penalty boundary. Yet a useful ranking is not a reliable audit
ledger. The preregistered gate addresses the latter; no alternative
endpoint-based admission threshold was validated here. Future evaluation
should specify both requirements in advance rather than choose a gate
after observing favorable endpoints.

\paragraph{Whose labels are right.}
A second alternative is that the human-gold arm's collateral and weaker
suppression reflect the gold rather than a property of human judgment. The
concentrated disagreements reported in Section~\ref{sec:res-e2e} permit
both annotation noise and a substantive difference in the evidentiary
standard. Re-adjudication needs the annotators' rationales and linked atoms,
not a preference for whichever source produces the cleaner ranking curve.
Until then, the adverse numbers remain part of the result under the current
gold version, without a claim that human judgment itself causes harm.

\subsection{The Threat Model Is Institutional, Not Hypothetical}

Fluent claims with absent or borrowed support recur in enforcement records
at commercial scale. Beyond the Truly Organic case discussed in the
introduction, examples include the USDA's list of fraudulent organic
certificates~\cite{usda_fraudulent_certs};
a GREENGUARD certificate absent from UL's registry, confirmed forged on
inquiry~\cite{leeduser2020greenguard}; and the FTC's first purchased-reviews
settlement (\$12.8M~\cite{ftc2019fakereviews}). The scarce resource is never
convincing text, which is cheap, but the evidence behind it: exactly what an
evidence-constrained ranker prices.

\subsection{Implications and Open Problems}

\paragraph{For platforms.}
Unsupported rank gain is a susceptibility, not a constant, so integrity
auditing must be per ranker: a clean result on one family certifies nothing
about another. The defense objective should shift from suppressing detail,
which taxes honest richness, to discounting unsupported query-relevant mass.
Every GroundedGEO adjustment carries a claim--evidence trace, providing
a basis for reviewing a demotion against an eligible source. A platform
would still need a procedure for contesting packet omissions: the existing
\texttt{unverifiable} state handles claims that cannot be adjudicated,
but does not automatically exempt adjudicable claims whose attestations
are missing.

\paragraph{Open problems.}
The next evaluation should first settle the disputed gold labels, then
test broader judge families against both ledger agreement and ranking
requirements on an expanded holdout. Separately, a retrieval intervention
is needed to test whether repairing missing evidence can recover the
protection lost in the thinning probe. Neither question can be resolved
by increasing the penalty strength alone.

\section{Limitations}
\label{sec:limitations}

\paragraph{Ranker and domain scope.}
The phenomenon is established on the tested rankers, with significant
unsupported-specific promotion on Qwen2.5-7B but weaker or absent evidence
on the other two models. This is not a universal attack claim. The ESCI
e-commerce setting~\cite{reddy2022shopping} also leaves transfer to local
services, B2B, and Web-passage ranking untested. A broader ranker-by-domain
evaluation is needed to map susceptibility rather than transfer the
reported magnitudes.

\paragraph{Supplied evidence and incomplete coverage.}
The closed-world estimand concerns support in a supplied packet, not
open-world truth (Section~\ref{sec:pf-scope}). The controlled thinning
probe establishes sensitivity to evidence loss; it does not estimate
how often a deployed retrieval system would omit a valid attestation.
Natural coverage in this benchmark is also uneven: the review layer covers
42 of 50 queries and the registry layer 18 credential queries. The
documented coverage split has eight uncovered queries, while a derived
split has nine (\texttt{q008} differs). Uncovered queries have higher
endpoints across arms, so they are not an exchangeable control group.
The headline estimates retain all 50 queries; coverage claims rely on
within-query thinning, pending reconciliation of the split and
per-query coverage reporting.

\paragraph{Human-gold size and unresolved judgments.}
Gold-dependent estimates rest on 370 claims from only 10 queries, not
370 independent ranking tasks. Broader certification requires a larger
adjudicated holdout. In addition, the current gold revision produces
clean-arm collateral and weaker attack suppression
(Table~\ref{tab:e2e-samecases}). Section~\ref{sec:res-e2e} identifies the
disputed base-identity and attack-arm labels; their causes remain
unresolved. The adverse results are retained, but neither annotation
noise nor a substantive difference in standards is established.
Re-adjudication with reason fields is necessary before treating this arm
as a settled reference. Zero FSR in the construction and local-judge
arms does not resolve that disagreement.

\paragraph{Operating-point identification.}
The development sweep does not establish a sharp optimum
(Section~\ref{sec:res-tuning}). Reported effects at $\lambda=40$ are
fixed-point comparisons, not a tuned maximum or validation of a frozen
deployment setting. Selecting a useful tradeoff requires an evaluation
regime with informative protected-arm errors, not merely a wider grid.

\paragraph{Judge and base-scorer breadth.}
The two local judges are checkpoints from one model family; the API
judge runs on the holdout only. Their gate failures do not establish
failure across judge families. All defense endpoints also use one
pointwise base scorer, DeepSeek-Flash. Sharing it controls the within-study
comparison, but neither absolute rates nor relative gains are established
for other scorers. Replication must vary the label source and base scorer
separately.

\paragraph{Target-only intervention and endpoint scope.}
Distractors carry no claims, so the penalty acts only on the target.
This limits conclusions about competition among multiple penalized
candidates and makes the target-specific normalization a special case
(Section~\ref{sec:discussion}). Moreover, the displayed defense tables
report raw top-3 rates and Success reduction, not the distinct frozen
clean-referenced Promotion reduction endpoint. They do not alone establish
that primary endpoint or the utility and grounding measures defined in
Section~\ref{sec:pf-estimands}.

\section{Ethical Considerations}
\label{sec:ethics}

This paper studies manipulation to defend ranking integrity. Attacks are
built on synthetic candidate descriptions in a controlled benchmark; no real
page, listing, or review was altered, and no live ranking system was
targeted. Enforcement cases in Section~\ref{sec:discussion} are public
records, not allegations. The work is dual-use: the same principles could
aid real manipulation. We disclose because fabricated query-aligned text is
already cheap---the scarce resource is evidence verification, which we
measure. We release the evaluation
apparatus---evidence-paired variants, packet twins, matcher reliability
gate---making misuse measurable.

Human annotation covered 370 product claims and contained no personal
information; annotator identities remain private. Two annotators labeled
each claim independently; automated suggestions are disclosed as such, never
as human judgments. Datasets are used under their research terms:
ESCI~\cite{reddy2022shopping}; Amazon Reviews 2023~\cite{hou2024bridging}.

Evidence-constrained ranking can disadvantage small publishers with thin
third-party coverage. The controlled coverage probe
(Section~\ref{sec:res-coverage}) demonstrates the mechanism of this harm,
not its distribution across real publishers. A deployment should expose
packet omissions, allow candidates to submit eligible attestations, and
provide review of disputed penalties. The claim--evidence trace supports
such review but is not itself an appeals process. Likewise, retaining
\texttt{unverifiable} avoids coercing unadjudicable claims into a verdict;
it does not protect every true claim absent from a packet. Source
eligibility should depend on independence rather than popularity, with
first-party documents used only for claims they can establish. These
governance measures require validation beyond the present experiments.

\section{Conclusion}
\label{sec:conclusion}

Evidence support is a relation between claims and records, not a property
of candidate prose. GroundedGEO makes that relation explicit at ranking
time through an evidence-paired benchmark, a claim-level reranker, and an
auditable decision trace. Fixed-text packet twins demonstrate a response
to evidence that a text-only rule cannot reproduce; matched-arm
experiments show that unsupported detail can receive ranking reward,
with susceptibility varying across the tested models.

The evidence channel offers conditional protection, not a completed
automatic defense. Construction labels yield selective suppression on
the supplied packets, and local automatic labels retain substantial
ranking effects despite failing the preregistered agreement gate.
Packet thinning exposes a separate vulnerability that better matching
alone cannot repair: honest claims lose protection when their evidence
is absent. Disputed human-gold labels and operating-point uncertainty
further limit the interpretation of the current endpoints.

The contribution is therefore an apparatus for separating ranking reward,
label reliability, and evidence availability, rather than treating a
plausible description or a favorable ranking curve as sufficient proof
of integrity. Progress requires validating the labels that justify a
penalty and the evidence collection that makes such justification possible.

\bibliographystyle{ACM-Reference-Format}
\bibliography{references}

\end{document}